**Self-injection locking of a vortex spin torque oscillator by delayed feedback**

S. Tsunegi[1,2], E. Grimaldi[1], R. Lebrun[1], H. Kubota[2], A.S. Jenkins[1], K. Yakushiji[2], A. Fukushima[2], P. Bortolotti[1], J. Grollier[1], S. Yuasa[2], and V. Cros[1]

[1]*Unité Mixte de Physique CNRS/Thales and Université Paris Sud, Palaiseau, France*
[2]*National Institute of Advanced Industrial Science and Technology (AIST), Spintronics Research Center, Tsukuba, Japan*

Abstract:

The self-synchronization of spin torque oscillators is investigated experimentally by re-injecting its radiofrequency (rf) current after a certain delay time. We demonstrate that the emission power and the spectral linewidth are improved for optimal delay times. Moreover by varying the phase difference between the emitted power and the re-injected one, we find a clear oscillatory dependence with a $2\pi$ periodicity of the frequency of the oscillator as well as its power and linewidth. Such periodical behavior within the self-injection regime is well described by the general model of nonlinear auto-oscillators including not only a delayed rf current but also all spin torque forces responsible for the self-synchronization. Our results reveal new approaches for controlling the non-autonomous dynamics of spin torque oscillators, a key issue for rf spintronics applications as well as for the development of neuro-inspired spin-torque oscillators based devices.

A major scientific breakthrough in spintronics was the introduction of spin transfer forces as a new mean to generate high frequency nonlinear dynamics in nanoscale magnetic devices. The wealth of physics in spin transfer phenomena paves the way to a new generation of multi-functional spintronic devices [1]. Recent trends range from nanoscale radiofrequency (rf) devices for efficient microwave source [2] to microwave detection [3,4], magnonic devices [5] and more recently neuro-inspired memory devices [6]. For the purpose of realizing these applications, it becomes of primal importance to not only identify and control the sources of noise and but also to achieve a fine control of the phase of these spin torque devices. Indeed, it is known and widely used in other types of oscillators such as conventional optical lasers [7] or voltage control oscillators [8] that the control of the oscillator phase can be achieved by a self-delayed feedback. In 2014, V, Tiberkevitch et al. [9] proposed a similar implementation for a spin-torque oscillator (STO) circuit based on delayed self-injection of the output rf current. It is to be noticed that the large nonlinear behavior, which is specific to these STOs might detrimentally impact the self-locking process of the device [10,11]. More recently, Khalsa *et al.* reported in a theoretical study that the control of linewidth reduction could be expected in a STO circuit based on delayed self-injection of the output rf current [12]. To our knowledge, this approach has not yet been addressed experimentally. We believe that a demonstration of the tuning of rf properties through a controlled delay would represent an important step for mastering the properties of STOs (frequency, spectral coherence and power consumption), which is crucial for the targeted rf applications [2] [13] as well for neuro inspired STO based memory devices [1,6].

Our main objective here is to identify the mechanisms of the self-injection locking of a vortex based STO using a delay line. In particular, we investigate the influence of the delay time $\Delta t$, on the main rf characteristics of this new oscillating regime. The studied samples are composed of circular FeB free layer based on magnetic tunnel junctions (MTJs). The complete stack of the MTJ consists of buffer/PtMn(15)/Co$_{70}$Fe$_{30}$(2.5)/Ru(1.0)/Co$_{60}$Fe$_{20}$B$_{20}$(2)/MgO(1.1)/Fe$_{80}$B$_{20}$(4.0)/MgO(1.1)/Ta(8) /Ru(7) where the subscript denotes the composition in atomic percent and the numbers in brackets indicate the layer thickness in nm (see ref [14]). Here, the top layer of the synthetic antiferromagnetic reference layer with uniform in-plane magnetization is the spin polarizing layer. The free layer made of FeB is covered with a MgO cap in order to decrease its magnetic damping that can be as small as 0.005 [15,16]. After annealing at 360°C in vacuum, magnetic tunnel junctions (MTJs) with radius of 150 nm were patterned by Ar ion milling and e–beam lithography. The typical magneto-resistance (MR) ratio is about 120% at room temperature and the MTJ resistance is around 53 Ω at bias voltage of 30 mV. For the FeB layer, the thickness and diameter were chosen so that the magnetic configuration at remanence corresponds to a magnetic vortex. All the measurements presented here have been carried out with magnetic field of $H_\perp =$ 3.0 kOe (the value necessary to have a large spin torque acting on the vortex [17]). Similar conclusions were obtained however for other $H_\perp$ values.

In Fig. 1b, we display a typical power spectral density (PSD) of the free-running STO i.e. without re-injection of the rf signal. The rf signal comes from the spin transfer induced sustained vortex oscillations, and exhibits a frequency of 316.6 MHz, an integrated power of 1.1 µW and a full-width at half-maximum (FWHM) of 310 kHz recorded under $I_{dc} = 4.0$ mA with a threshold current being $I_{dc} = 2.1$ mA. In order to re-inject the rf signal generated by the STO into the oscillator, we use the measurement circuit described in Fig. 1a. The generated rf signal passes through a bias-tee, rf cables and eventually through the input port of a directional coupler. A close end at the output port of the directional coupler permits the reflection of the rf signal and injects the signal back into the STO with an intensity close to about 40% of the generated rf power. Note that most of the losses are from the cables. A tunable delay line is inserted in the circuit in order to precisely control the phase difference between the STO and the re-injected rf current which is defined as: $\Delta\theta = 2\pi f_{sto} \Delta t + \pi$ where $f_{sto}$ is the STO frequency with re-injection and $\Delta t$ is the total delay time introduced by the circuit. This delay time $\Delta t$ comprises the delay due to the rf components and the delay due to the cables measured independently using a vector network analyzer (VNA). The last term $\pi$ is added because of the phase shift which occurs at the close end. In the measurements presented here, the good impedance matching of the MTJ allows us to disregard the presence of stationary waves in this circuit (see Supplementary Materials). The coupled port of the directional coupler is used to measure the resulting rf signal generated from the STO after re-injection.

In Figs. 1c and d, we present PSD curves measured at $I_{dc}$ = 4.0 mA when the rf signal is re-injected into the STO with two different delay times $\Delta t$ (obtained by adjusting the length of the tunable delay line). For $\Delta t$ = 37.6 ns (shown in Fig 1c), we find that the frequency decreases down to 314.5 MHz i.e. 2.1 MHz lower than the free-running case. At the same time, the emission power decreases to 1.02 µW. When the delay is tuned to $\Delta t$ = 38.6 ns (see Fig 1d), the frequency becomes 318.6 MHz and the emission power increases up to 1.18 µW which is the highest value that can be obtained by varying the delay time at $I_{dc}$ = 4.0 mA. We also measure the PSDs obtained for longer delay time $\Delta t$ (in other words, a larger phase difference $\Delta\theta$). With these additional measurements (see Fig. 1e), we clearly observe a sinusoidal $2\pi$–dependence of the STO peaks on delay time $\Delta t$. To our knowledge, such oscillating dependence on $\Delta\theta$ represents the first experimental demonstration of the self-injection locking of STO on its own rf emitted current.

In the following, we focus on measurements of self-injection locking performed under the condition $I_{dc}$ = 3.7 mA, the condition at which the STO presents a relatively large nonlinear parameter $\nu$ of 4.1 as deduced from phase and amplitude noise analysis [18,19]. In Fig. 2a, we show a clear $2\pi$- dependence of the normalized power $p_0$ (calculated from the square of the oscillation amplitude of vortex core) that varies between 0.255 and 0.295. As for the STO frequency $f_{sto}$ with the phase difference $\Delta\theta$ (see Fig. 2b), we find that its variation in the region between $\Delta\theta = 0$ and $\Delta\theta = 5\pi$ is around 0.8%, equivalent to 2.6 MHz of the value measured without re-injection (see dotted line in Fig. 2b). These experimental results clearly indicate that the re-injected rf current significantly modifies the limit cycle of the oscillating vortex core and defines a new oscillating regime. To quantify the amplitude of the rf re-injected current, we performed measurements using a VNA and found the amplitude to be about 80 µA i.e. about 2% of the dc current. We also stress that the self-synchronization has been achieved without any amplification of the rf current emitted by the STO.

To understand the main features of the mechanisms of self-injection locking of STO using delayed feedback, we refer to the analytical analysis of this system recently done by Khalsa *et al.* [15]. Rewriting Eq. (5) and (8) of Ref. [15] using the more conventional notations of the nonlinear auto-oscillator theory proposed by Slavin and Tiberkevitch [20] gives:

$$p_0 = p_0^* \left\{ 1 - \frac{F}{\Gamma_p^*} \cos(\Delta\theta + \varphi_{STT}) \right\} \quad (4a)$$

$$f_{STO} = f_{STO}^* + \frac{\sqrt{1+\nu^2} F}{2\pi} \sin(\Delta\theta + \varphi_{STT} - \tan^{-1}(\nu)) \quad (4b)$$

In these equations, $f_{STO}^* = \frac{\kappa}{2\pi G}(1 + \zeta p_0^*)$, $p_0^* = \frac{Ga_J J_{dc} - D\kappa}{D\kappa(\xi+\zeta)}$ and $\Gamma_p^* = \frac{Ga_J J_{dc} - D\kappa}{G^2}$ are respectively the frequency, the normalized power and the relaxation damping rate of the stationary free-running STO. Several coefficients govern the dynamics of the oscillator: the vortex gyrovector $G$, the linear damping $D$, the nonlinear damping $\xi$, the linear confinement stiffness $\kappa$, the nonlinear confinement $\zeta$ and the Slonczewski torque efficiency $a_J$ associated with the perpendicular component of the spin polarization and responsible for the free-running spin transfer induced vortex oscillation. The amplitudes of power and frequency variations depend on the strength of the normalized self-synchronization force $F$, expressed as $F = \sqrt{\Lambda_{SL//}^2 + \Lambda_{FL//}^2} C_{MR} J_{dc}/2G$ where $C_{MR}$ is a proportionality factor including the circuit losses and the MR ratio of the MTJ. $F$ depends on the two spin torques capable of driving the vortex synchronization: the field like torque $\Lambda_{FL//}$ and the Slonczweski torque $\Lambda_{SL//}$ originating from the in-plane spin polarization. Both normalized power and frequency are expected to evolve as a sine function of the phase difference between the emitted and re-injected signal $\Delta\theta = \theta(t) - \theta(t - \Delta t)$. However, they are dephased when compared with $\Delta\theta$. This phase shift $\varphi_{STT} = \tan^{-1}(\frac{\Lambda_{FL//}}{\Lambda_{SL//}})$ depends on the one which is the larger of the two spin torques. In addition, the frequency phase shift depends also on the nonlinearity of the vortex $\nu$.

We now compare these analytical predictions to our experimental results. As shown in Fig.2, both the normalized power $p_0$ and frequency $f_{STO}$ follow a sinusoidal dependence on $\Delta\theta$ in the self-synchronized regime, in agreement with the predictions of Eq. 4. We first discuss the amplitude of power and frequency variations with $\Delta\theta$. Equation 4a indicates that the power $p_0$ should be inversely proportional to the relaxation damping rate $\Gamma_p^*$. As detailed in the Supplementary Materials, we have been able to confirm this dependence with $\Gamma_p^*$, which further validates the self-injection locking model of Eq. 4. The changes of frequency $f_{STO}$ should be directly linked to the non-linear parameter $\nu$ as expected from the prefactor of the sine in Eq.4b. In Fig.2, we find a frequency variation with $\Delta\theta$ as large as 2.6 MHz when $\nu = 4.1$. For the measurements shown in Fig. 1e, a smaller variation amplitude (about 2.2 MHz) is obtained in agreement with a smaller $\nu = 3.1$ at $I_{dc} = 4.0$ mA. A more complete study can be found in the Supplementary Materials which also confirms the validity of the model. We now focus on the observed phase shifts of frequency and power. We first emphasize that in Fig.2, $f_{STO}$ and $p_0$ are oscillating almost in phase, with a very small phase difference of about $0.05\pi$. This behavior is expected in highly non-linear oscillators. Indeed in Eq. 4 the term $\tan^{-1}(\nu)$ factor is always close to $\pi/2$ as long as the $\nu$ parameter is larger than 3 which is the case for all these measurements.

Using Eq.(4a) and Eq. (4b) and having evaluated the $\nu$ parameter, we can estimate the spin-transfer-forces phase shift $\varphi_{STT}$ based on the dependence of $p_0$ and $f_{STO}$ on $\Delta\theta$ (see Fig. 2a and 2b). Both dependencies result in a very similar $\varphi_{STT}$ value, around $1.4\pi$ for $I_{dc} = 3.7$ mA. It has to be noticed that a value close to $3\pi/2$, implies that the field-like-torque drives the synchronization in our FeB MTJs i.e. $\Lambda_{FL//} \gg \Lambda_{SL//}$. This specific feature of vortex based STO is important as, in general, the synchronization mechanisms equally depend on both $\Lambda_{SL//}$ and $\Lambda_{FL//}$ and on their sign. We have repeated the same analysis for different dc currents and have extracted the $\varphi_{STT}$ dependence on $I_{dc}$ (see Fig.3). The evolution in the whole current range (between 3.0 and 4.5 mA) shows that $\varphi_{STT}$ only slightly increases with $I_{dc}$, presumably because of the different bias voltage dependences of the two torques [21].

Another important parameter of spin transfer induced oscillations is the threshold dc current $J_c$ for sustained oscillations in the self-synchronized regime. In Fig. 4, we display the experimental threshold current $J_c$ dependence on $\Delta\theta$ that has been estimated from the inverse power $p_0$ dependence on $J_{dc}$ for different values of $\Delta\theta$. We find that $J_c$ displays a clear periodic behavior with $\Delta\theta$ in agreement with Eq. (4c) and can thus be reduced for optimal delays. For particular values of the delay time, $J_c$ is therefore decreased, which provides an interesting route to explore towards spin torque oscillations with reduced power consumption. This $2\pi$-periodic evolution with $\Delta\theta$ of the critical current $J_c$ is also in agreement with the analytical model [15]:

$$J_C \approx \frac{J_C^*}{1 - \frac{GF}{a_J J_{dc}} \cos(\Delta\theta + \varphi_{STT})} \quad (4c)$$

Note that in Fig.4, the $\varphi_{STT}$ values extracted from the analytical expression of $J_C$ in Eq. 4c is again $1.5\pi$, which is in excellent agreement with the one previously extracted from the $p_0$ and $f_{STO}$ evolutions shown in Fig. 3.

We now focus on the impact of the self-injection process on the spectral quality of the STO. In Fig. 5, we display the evolution of the experimental linewidth (see blue dots) as a function of the phase difference $\Delta\theta$ measured in the self-synchronized regime. Based on this mechanism related to the use of a delay, we demonstrate that the STO linewidth can be reduced from 470 kHz in the free running regime down to 180 kHz in self-synchronized regime. This result clearly highlight the advantage of using a delay line from an application point of view as it permits to optimize the linewidth of the vortex STO with an optimum phase shift $\Delta\theta$ and a large delay time. In order to unravel the mechanisms responsible for the experimentally observed variation of the linewidth, we again compare our experimental results with analytical predictions. Khalsa *et al.* calculated that (for linewidth smaller than the typical relaxation rate $\Gamma_p$), the linewidth of the self-synchronized regime can be expressed as :

$$FWHM = \frac{2\Delta f_0 \,(1 + \nu^2)}{\left[1 - F\Delta t \sqrt{1 + \nu^2} \, \sin(\Delta\theta + \varphi_{stt} + \arctan(1/\nu))\right]^2} \quad (5)$$

where $2\Delta f_0$ is the linear linewidth associated with the self-synchronized stationary power $p_0$.

By calculating the amplitude noise auto-correlation function of the self-synchronized STO (see Supplementary Material), we can extend this result and rewrite it in a more concise and physical manner as :

$$\text{FWHM} \sim = 2\Delta f_0 \, (1 + \nu^2)/\lambda^2 \quad (6)$$

In this equation, $\lambda$ is directly the factor renormalizing the relaxation damping rate in the self-synchronized regime: $\lambda\Gamma_{p \text{ self-sync}} = \lambda\Gamma_p$. Analyzing the different terms in Eq. 6, we notice that the delay $\Delta t$ can influence the STO linewidth through two different mechanisms. The first mechanism is indirect. Indeed, the linear linewidth $\Delta f_0$ depends inversely on the power $p_0$[19,20], which oscillates with $\Delta\theta$ as we have seen previously. If this mechanism is the main process for the linewidth evolution, then we expect to get linewidth maxima (resp. minima) for stationary power $p_0$ minima (resp. maxima). In Fig. 5, we display the expected oscillating behavior of linewidth with $\Delta\theta$ due to the change of the stationary regime i.e. only taking into account the numerator $2\Delta f_0 (1+\nu^2)$ (see green curve). We clearly see that the two curves show distinct behavior, thus discarding this mechanism of linewidth evolution with delay. The second mechanism which can lead to a change of linewidth is related to the factor $\lambda$ renormalizing the relaxation damping rate and thus corresponds to the intrinsic noise filtering associated with the length of delay. In Fig. 5, we also plot the predicted evolution of $2\Delta f_0(1+\nu^2)/\lambda^2$ with $\Delta\theta$ and clearly see a very good qualitative agreements with the experimental results, notably on the position of maxima and minima with $\Delta\theta$. This result shows that the measured large variation of linewidths induced by the delayed feedback is directly due to the modified phase and amplitude dynamics in the self-synchronized regime.

In conclusion, the self-synchronization of STO has been successfully demonstrated for the first time by using a delayed feedback circuit. The self-synchronization induces new stationary regimes and endows the STO parameters with a periodic behavior. When the phase difference is appropriately tuned by optimizing the delay time, we find that the STO spectral linewidth can be significantly reduced (more than 60% of reduction compared with the free running STO) and the emitted power increased compared with their respective values without self-synchronization. Such periodical behavior within the self-injection regime is well explained by considering the large field-like spin transfer force. Our results achieve new approaches for controlling the phase of spin torque oscillators, a key issue for rf spintronics applications as well as for the development of neuro-inspired STO based devices.

Acknowledgements

The authors acknowledge Jacob Torrejon Sanchez and Matthieu Riou for fruitful discussion. The financial support from JSPS KAKENHI Grant Number 23226001, from EU grant (MOSAIC No. ICT-FP7- n.317950) and from ANR agency (SPINNOVA) is acknowledged. E.G. thanks DGA and CNES for supporting her PhD fellowship.
*Correspondence should be addressed to S. Tsunegi (tsunegi.sb@aist.go.jp) and V. Cros (vincent.cros@thalesgroup.com)

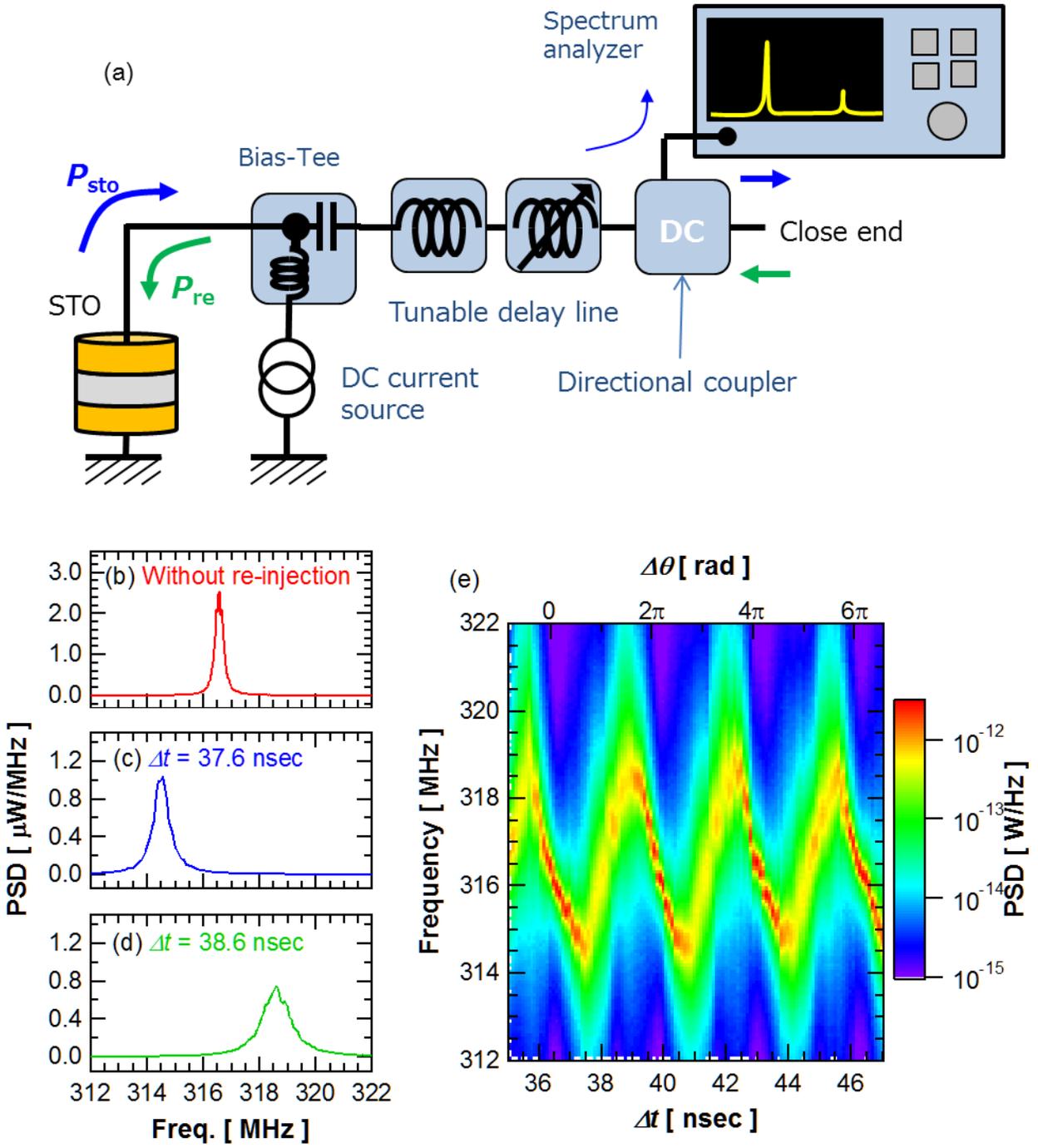

Fig. 1 : (a) Schematic of the delayed feedback circuit. Power Spectral Density (PSD) spectra at $I_{dc}$=4.0 mA (b) without re-injection, with re-injection (c) $\Delta t$ = 37.6 ns, and (d) $\Delta t$ = 38.6 ns. (d) Image plot of the PSD spectra at $I_{dc}$ = 4.0 mA.

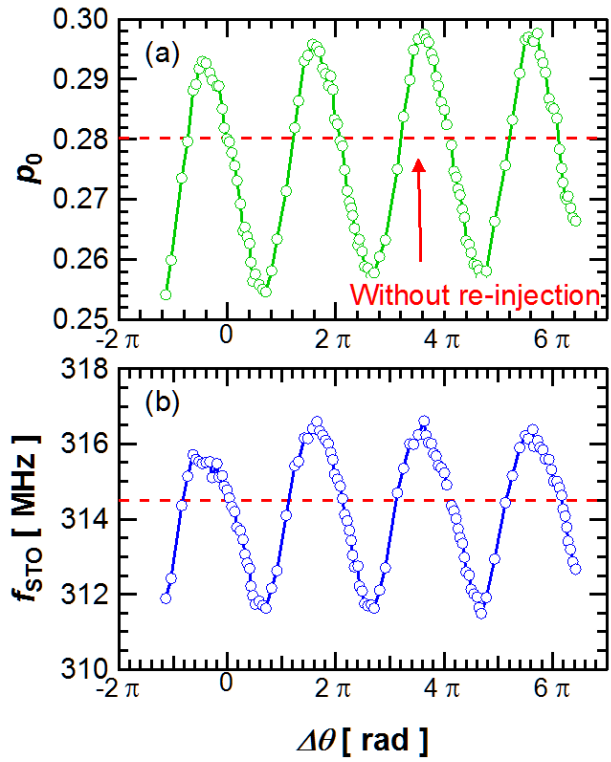

Fig.2 : Measurements (a) the normalized power $p_0$ and (b) the STO frequency $f_{STO}$ evolution as a function of the phase difference $\Delta\theta$ of at $I_{dc}$ = 3.7 mA. The dotted red line is the value measured without re-injection (free-running STO) for the same external conditions.

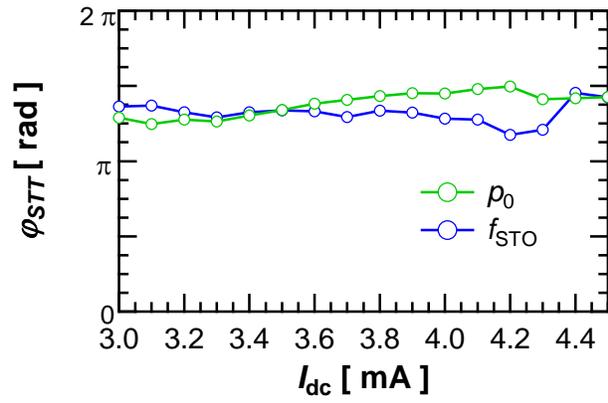

Fig.3 : Evolution of the estimated phase shift $\varphi_{STT}$ with the dc current $I_{dc}$

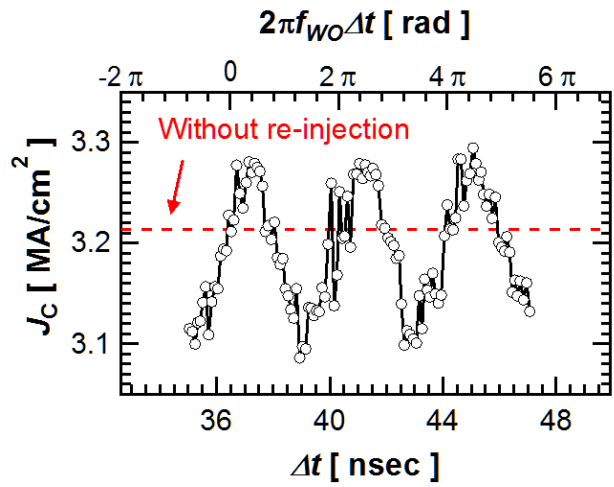

Fig.4 : Critical current density $J_C$ dependence on phase difference. The dotted red line is the value of the critical current density for that without re-injection (free-running STO).

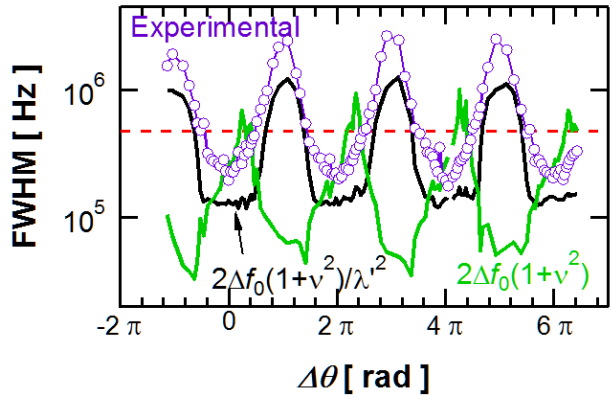

Fig.5 : Evolution of the experimental spectral linewidth (blue curve) with the phase difference $\Delta\theta$ at $I_{dc}$ = 3.7 mA. The dotted red line is the value of the FWHM for that without re-injection (free-running STO). The black curve corresponds to the predicted linewidth evolution obtained from Eq.5 in the main text. The green curve describes the modification of the linewidth due only to a change of stationary regime.